\documentclass[twocolumn,preprintnumbers,superscriptaddress,prd,nofootinbib]{revtex4-1}
\usepackage{graphicx}
\usepackage{times}
\usepackage{epsfig}
\usepackage{epstopdf}
\usepackage{amsmath}
\usepackage{amsfonts}
\usepackage{amssymb}
\usepackage{url}
\usepackage{subfigure}
\usepackage{verbatim}
\usepackage{comment}

\newcommand{\be}{\begin{equation}}
\newcommand{\ee}{\end{equation}}
\newcommand{\bea}{\begin{eqnarray}}
\newcommand{\eea}{\end{eqnarray}}

\begin{document}

%\preprint{}
% Page numbers bottom-center
\pagestyle{plain}
%%%%%%%%%%%%%%%%%%%%%%%%%%%%%%%%%%%%%%%%%%%%%%%%%%%%%%%%%%%%
\title{
Inflection-Point Inflation with Axion Dark Matter \\
 in light of Trans-Planckian Censorship Conjecture
}
%%%%%%%%%%%%%%%%%%%%%%%%%%%%%%%%%%%%%%%%%%%%%%%%%%%%%%%%%%%%%
%\author{Nobuchika Okada}
%\email{okadan@ua.edu}
%\affiliation{
%Department of Physics and Astronomy,
%University of Alabama,
%Tuscaloosa, AL 35487, USA
%}
%\author{Qaisar Shafi}
%\Email{shafi@bartol.udel.edu}
%\affiliation{
%Bartol Research Institute,
%Department of Physics and Astronomy,
%University of Delaware, Newark, DE 19716, USA
%}
%
\author{Nobuchika Okada}
\email{okadan@ua.edu}
\affiliation{
Department of Physics and Astronomy,
University of Alabama,
Tuscaloosa, AL 35487, USA
}
\author{Digesh Raut}
\email{draut@udel.edu}
\affiliation{
Bartol Research Institute,
Department of Physics and Astronomy,
University of Delaware, Newark, DE 19716, USA
}
\author{Qaisar Shafi}
\email{qshafi@udel.edu}
\affiliation{
Bartol Research Institute,
Department of Physics and Astronomy,
University of Delaware, Newark, DE 19716, USA
}

%\date{\today}

%\baselineskip 36pt

%%%%%%%%%%%%%%%%%%%%%%%%%%%%%%%%%%%%%%%%%%%%%%%%%%%%%%%%%%
\begin{abstract}

Motivated by the recently proposed Trans-Planckian Censorship Conjecture (TCC), we propose a gauged $B-L$ model
of inflection-point inflation with axion dark matter. 
The Hubble scale during inflation ($H_{\rm inf}$) satisfies the TCC bound of $H_{\rm inf} \lesssim 1$ GeV, 
the axion dark matter scenario is free from the axion domain wall and isocurvature problems. 
The seesaw mechanism is automatically incorporated in the model and 
the observed baryon asymmetry of the universe can be reproduced via resonant leptogenesis. 

\end{abstract}
\maketitle

%%%%%%%%%%%%%%%%%%%%%%%%    
\section{Introduction} 
%%%%%%%%%%%%%%%%%%%%%%%%

The recently proposed Trans-Planckian Censorship Conjecture (TCC) \cite{Bedroya:2019snp} effectively states that
  sub-Planckian quantum fluctuations in an expanding universe should remain quantum
  and never get larger than the Hubble horizon.  
Applied to inflationary cosmology this conjecture yields an approximate upper bound of around $10^9$ GeV
  on the energy scale during inflation, or equivalently, the upper bound on the Hubble scale $H_{\rm inf} \lesssim 1$ GeV
  during the (slow-roll) inflation \cite{Bedroya:2019tba}. 
The TCC offers a potential resolution of the so-called Trans-Plankian problem \cite{Brandenberger:2000wr}
  for a low energy effective theory to be consistent 
  with quantum gravity and avoid being relegated to the ``swampland" \cite{Obied:2018sgi} (see, also, Ref.~\cite{ Dvali:2018fqu}).  
Clearly, the TCC conjecture favors low scale inflation models, which motivates us here
   to implement inflection-point inflation \cite{Okada:2016ssd} in a gauged U(1)$_{B-L}$ extension of the Standard Model (SM). 
(For recent work on inflation scenarios compatible to the TCC, see Ref.~\cite{recent}).     
To account for the missing dark matter (DM) in the SM, we include a U(1) Peccei-Quinn (PQ) symmetry \cite{Peccei:1977hh}, 
   which resolves the strong CP problem of the SM and also provides a compelling DM candidate
   in the form of axion \cite{Weinberg:1977ma, Wilczek:1977pj}.     
In order to cancel gauge anomalies due to the presence of the $B-L$ symmetry, 
   three right-handed neutrinos have to be introduced. 
The presence of these latter neutrinos allows one to account for the observed solar and atmospheric neutrino oscillations.
Furthermore, the observed baryon asymmetry can be explained via leptogenesis \cite{Fukugita:1986hr}.

This brief note is organized as follows: 
In the next section, we propose a U(1)$_{B-L} \times$U(1)$_{PQ}$ extension of the SM, 
  which appends U(1)$_{PQ}$ to the minimal $B-L$ model \cite{mBL}. 
In Sec.~\ref{sec:3}, we discuss inflection-point inflation with $H_{\rm inf} \lesssim 1$ GeV 
  to satisfy the TCC bound, and in Sec.~\ref{sec:4}, we discus reheating after inflation and leptogenesis. 
An axion DM scenario is discussed in Sec.~\ref{sec:5}, and our conclusions are presented in Sec.~\ref{sec:6}.

%%%%%%%%%%%%%%%%%%%%%%%%%%%%%%%
\section{U(1)$_{B-L} \times$U(1)$_{PQ}$ extension of the SM}
\label{sec:2}
%%%%%%%%%%%%%%%%%%%%%%%%%%%%%%%
%%%%%%%%%%%%%%%%%%%%
\begin{table}[t]
\begin{center}
\begin{tabular}{|c|ccc|c|c|c|}
\hline
                       &SU(3)$_C$        &SU(2)$_L$       &U(1)$_Y$          & U(1)$_{B-L}$                        & U(1)$_{PQ}$
\\ 
\hline
$q_L^{i}$           &{\bf 3}             &{\bf 2}            &$ 1/6$          & $ 1/3$         & $ 1  $   
\\
$u_R^{i}$     &{\bf 3}     &{\bf 1}           &$2/3$ & $ 1/3$       & $ -1 $   
\\
$d_R^{i}$     & {\bf 3}    &{\bf 1}           &$-1/3$          & $ 1/3$      & $ -1$   
\\
\hline
$\ell_L^{i}$         &{\bf 1}            &{\bf 2}            &$-1/2$              & $ -1$           &$ 1$
\\
$e_R^{i}$      &{\bf 1}            &{\bf 1}           &$ -1$              & $-1$           &$ -1$   
\\
$N_R^{i}$   &{\bf 1}           &{\bf 1}            &$ 0$             & $ -1$                                & $-1$   
\\
\hline
$H_u$                & {\bf 1 }          &{\bf 2}            &$ 1/2$              & $ 0$                    & $-2$   
\\  
$H_d$                & {\bf 1 }          &{\bf 2}            &$- 1/2$            & $ 0$                     & $-2$   
\\
$S$                    & {\bf 1 }           &{\bf 1}            &$ 0$            &$ 0$                                  & $+4$
\\  
$\Phi$                &{\bf 1 }           &{\bf 1}            &$ 0$             &$+2$                                  & $+2$     
\\
\hline
\end{tabular}
\end{center}
\caption{
Particle content of the model. 
In addition to the three generations of SM fermions ($i=1,2,3$), three RHNs ($N_R^{i}$) are introduced. 
The scalar sector has two Higgs doublet fields ($H_{u}$ and $H_d$) and two SM singlet Higgs fields ($\Phi$ and $S$). 
}
\label{tab:1}
\end{table}
%%%%%%%%%%%%%%%%%%%%%%%%%%%%%%%%%%%%%%%%%%%%%%%

The minimal $B-L$ model \cite{mBL} is a well-motivated  extension of the SM, 
  where the global $B-L$ (baryon minus lepton number) symmetry of the SM is gauged. 
The introduction of three right-handed neutrinos (RHNs) is crucial to cancel all the gauge and
  mixed gauge-gravitational anomalies. 
The RHNs acquire Majorana masses from the U(1)$_{B-L}$ symmetry breaking, 
  and after the electroweak symmetry breaking, the type-I seesaw mechanism \cite{Seesaw} generates tiny masses for the observed neutrinos.   
The particle content of our U(1)$_{B-L} \times$U(1)$_{PQ}$ model is listed in Table~\ref{tab:1}. 
The SM singlet scalar field $\Phi$ breaks the $B-L$ symmetry and is crucial for implementing the inflection-point inflation. 
The introduction of two SM Higgs doublets ($H_{u,d}$) and a U(1)$_{B-L}$ singlet scalar ($S$) 
  is crucial to incorporate the PQ symmetry. 
One may regard our model as a U(1)$_{B-L}$ extension of 
  the Dine-Fischler-Srednicki-Zhitnitsky (DFSZ) model \cite{Dine:1981rt, Zhitnitsky:1980tq} 
  with $S$ being the PQ field.

The gauge and U(1)$_{PQ}$ invariant Higgs potential is given by  
\bea  
V &=& 
 - \sum_{i = u,d} \mu_i^2 \left( H_i^{\dagger}H_i \right)
+\sum_{i = u,d}\lambda_i  \left( H_i^{\dagger}H_i \right)^2 
\nonumber \\
&&
+  \left( \sqrt{2}\Lambda_{s}\left( H_u \cdot H_d \right) S + {\rm h. c.}\right)
\nonumber \\
&&
+\lambda_{\phi} \left(\Phi^\dagger \Phi  - \frac{v_{BL}^2}{2}\right)^2 
+\lambda_{S} \left(S^\dagger S  - \frac{v_{PQ}^2}{2}\right)^2 
\nonumber \\
&&+ {\rm mixed \; quartic \; terms}, 
\label{eq:HPot}
\eea
where $\Lambda_s$, $v_{BL}$ and $v_{PQ}$ are mass parameters, 
  all couplings are chosen to be real and positive,    
  the {\it dot} represents contraction of $SU(2)$ indices by $\epsilon$ tensor, 
  and the last term on the right-hand side indicates the mixed quartic scalar couplings
  such as $(H_i^\dagger H_i)(\Phi^\dagger \Phi)$, $(H_i^\dagger H_i)(S^\dagger S)$, etc. 
For simplicity, we assume that the mixed quartic couplings involving $\Phi$ and $S$ are negligibly small, 
  and the $B-L$ and the PQ symmetry breaking scales are much higher than 
  the electroweak scale and $\Lambda_s$.

With the vacuum expectation values (VEVs) of $\Phi$ and $S$ given by
  $ \langle \Phi \rangle =  v_{BL}/\sqrt{2}$ and $\langle S \rangle =  v_{PQ}/\sqrt{2}$, respectively,   
 the SM singlet Higgs fields can be parameterized as 
\bea
\Phi(x) &=& \frac{1}{\sqrt{2}}\left(\phi(x) + v_{X}\right) e^{i \chi(x)/ v_{X}}, 
\nonumber \\
S(x) &=& \frac{1}{\sqrt{2}}\left(s(x) + v_{PQ}\right) e^{i a(x)/ v_{PQ}}. 
\label{eq:phiABH} 
\eea 
The field $\chi$ is the would-be Nambu-Goldstone (NG) boson that is absorbed by the U(1)$_{B-L}$ gauge boson ($Z^\prime$), 
and the field $a(x)$ is the NG boson (axion) associated with the PQ symmetry breaking. 
The masses for $\phi$ and $s$, and the $Z^\prime$ boson are given by
\bea 
m_\phi = \sqrt{2 \lambda_\phi} v_{BL}, \; 
m_s = \sqrt{2 \lambda_s} v_{PQ},  \;
m_{Z^\prime} = 2 g v_{BL},
\label{eq:mass}
\eea
respectively, where $g$ is the $B-L$ gauge coupling.

With our assumption of negligible mixed quartic couplings, we analyze the Higgs doublets sector 
  separately from the $\Phi$ and $S$ sectors. 
The PQ symmetry breaking generates a mixing mass term for the two Higgs doublets, 
  $m_3^2 (H_u \cdot H_d)$, where $m_3^2=\Lambda_s v_{PQ}$.  
Note that with the PQ charge assignments listed in Table \ref{tab:1}, 
  $H_u$ ($H_d$) only couples with the up-type (down-type) SM fermions. 
Therefore, the low energy effective theory of our model after the $B-L$ and PQ symmetry breakings  
  is nothing but the type-II two Higgs doublet SM (+ axion). 
Since this two Higgs doublets model is well studied (see, for example, \cite{Branco:2011iw}), 
  we skip the detailed phenomenology of the low energy effective theory.

In addition to the Yukawa interactions of the SM leptons and quarks, 
  the following new Yukawa interactions involving the RHNs are introduced: 
\bea
\mathcal{L} _{Y} \supset - \sum_{i, j=1}^{3}  Y_{D}^{ij}  \overline{N_{R}^{j}}  (\ell_{L}^{i} \cdot H_u) 
 - \sum_{k=1}^{3} \frac{1}{2} Y_k  \Phi \overline{N_{R}^{k C}} N_{R}^{k} ,
\label{U1XYukawa}
\eea 
where $Y_D$ ($Y_k$) is the Dirac (Majorana) neutrino Yukawa coupling, 
   and we have chosen a flavor-diagonal basis for the Majorana Yukawa coupling without loss of generality.    
Through the $B-L$ and the electroweak symmetry breakings, 
  the Dirac and the Majorana masses for the neutrinos are generated: 
\bea 
m_D^{ij}= \frac{Y_D^{ij}}{\sqrt{2}} v_{u}, \; \;\;\;\;
m_{N^i}= \frac{Y_i}{\sqrt{2}} v_{BL},   
\label{eq:masses}
\eea
where $v_u$ is the VEV of $H_u$.

%%%%%%%%%%%%%%%%%%%%%%%%    
\section{Inflection-point inflation}
\label{sec:3}
%%%%%%%%%%%%%%%%%%%%%%%%
The inflection-point inflation (IPI) scenario is a unique low scale inflation scenario
  driven by a single scalar field $\phi$ \cite{Ballesteros:2015noa, Choi:2016eif, Okada:2016ssd}.
We begin by highlighting the key results of the IPI scenario. 
[See Ref.~\cite{Okada:2016ssd} for details.]  
Assuming that the inflaton potential exhibits an (approximate) inflection point, 
  we express the inflaton potential around $\phi = M$ at the approximate inflection point as   
\bea
V(\phi)\simeq V_0 +\sum_{n = 1}^3 \frac{1}{n!}V_n (\phi-M)^n , 
\label{eq:PExp}
\eea
   where $V_0 = V(M)$, $V_n \equiv  d^{n}V/ d \phi^n |_{\phi =M}$, 
   and the point $\phi = M$ is identified with the inflaton value at the horizon exit
   corresponding to the pivot scale $k_0= 0.05$ Mpc$^{-1}$
   of the Planck 2018 measurements \cite{Akrami:2018odb}.

With the inflaton potential in Eq.~(\ref{eq:PExp}), the slow-roll parameters are expressed as 
\bea
  \epsilon \simeq \frac{M_P^2}{2} \left( \frac{V_1}{V_0}\right)^2, \; 
  \eta \simeq M_P^2 \left( \frac{V_2}{V_0}\right), \; 
  \zeta \simeq M_P^4 \frac{V_1 V_3}{V_0^2} ,
\eea
where $M_P = 2.43\times 10^{18}$ GeV is the reduced Planck mass. 
The inflationary predictions are given by 
\bea
n_s=1- 6\epsilon +2 \eta, \; 
r=16 \epsilon, \;
\alpha= 16 \epsilon \eta -24 \epsilon^2 -2 \zeta^2, 
\eea     
where $n_s$, $r$ and $\alpha$ are the scalar spectral index, the tensor-to-scalar ratio, and the running of the spectral index, respectively.
The amplitude of the curvature perturbation $\Delta_{\mathcal{R}}^2$ is given by  
\bea
  \Delta_{\mathcal{R}}^2 \simeq \frac{V_0}{24 \pi^2 M_P^4 \epsilon}. 
\eea

To be consistent with the central values from the Planck 2018 results \cite{Akrami:2018odb},  
  $\Delta_{\mathcal{R}}^2= 2.099 \times 10^{-9}$  and $n_s = 0.965$, 
$V_{1,2}$ are expressed as  
\bea
\frac{V_1}{M^3}&\simeq& 2.01 \times 10^3 \left(\frac{M}{M_P}\right)^3\left(\frac{V_0}{M^4}\right)^{3/2}, \nonumber \\
\frac{V_2}{M^2}&\simeq& -1.73 \times 10^{-2}  \left(\frac{M}{M_P} \right)^2 \left(\frac{V_0}{M^4}\right).  
\label{eq:FEq-V1V2}
\eea
The number of e-folds ($N$) for the IPI scenario is approximated as \cite{Okada:2016ssd}
\bea
N\simeq \pi \frac{V_0}{M_{P}^2\sqrt{2 V_1 V_3}}, 
\label{eq:efold} 
\eea
which together with Eq.~(\ref{eq:FEq-V1V2}) leads to \footnote{Using Eqs.~(\ref{eq:FEq-V1V2}) and (\ref{eq:FEq-V3}), we can verify that the inflaton potential in Eq.~(\ref{eq:PExp}) has no exact inflection-point for our choice of $N= 40$.} 
\bea
\frac{V_3}{M} &\simeq& 1.54 \times 10^{-7} \; \left( \frac{40}{N} \right)^2
   \left( \frac{M}{M_P} \right) \left( \frac{V_0}{M^4}   \right)^{1/2}. 
\label{eq:FEq-V3}
\eea

%%%%%%%%%%%%%%%
\begin{comment}    
To be consistent with the central values from the Planck 2018 results \cite{Akrami:2018odb},  
  $\Delta_{\mathcal{R}}^2= 2.099 \times 10^{-9}$  and $n_s = 0.965$, 
$V_{1,2,3}$ are expressed as  
\bea
\frac{V_1}{M^3}&\simeq& 2.01 \times 10^3 \left(\frac{M}{M_P}\right)^3\left(\frac{V_0}{M^4}\right)^{3/2}, \nonumber \\
\frac{V_2}{M^2}&\simeq& -1.73 \times 10^{-2}  \left(\frac{M}{M_P} \right)^2 \left(\frac{V_0}{M^4}\right), \nonumber \\
\frac{V_3}{M} &\simeq& 6.83 \times 10^{-7} \; \left( \frac{60}{N} \right)^2
   \left( \frac{M}{M_P} \right) \left( \frac{V_0}{M^4}   \right)^{1/2}, 
\label{eq:FEq-V1V2}
\eea
where $N$ is the number of e-folds which is defined as 
\bea 
N = \frac{1}{M_P^2} \int_{\phi_e}^M \, d\phi \, \frac{V}{(dV/d\phi)}
\eea
with $\phi_e$ is the inflaton field value at the end of inflation determined by $\epsilon(\phi_e)=1$. 
\end{comment}    
%%%%%%%%%%%%%%%%

In order to solve the horizon problem of big bang cosmology, 
  the number of e-folds is expressed as
\bea
N \simeq 42.0 +\frac{1}{3} {\rm ln}\left[\frac{H_{\rm inf}}{0.1\, {\rm GeV}}\right] 
+\frac{1}{3} {\rm ln}\left[\frac{T_R}{10^{7} {\rm GeV}}\right], 
\label{eq:efolds}
\eea
%where we have used the Hubble relation during the inflation 
%$H_{inf}^2 \simeq (1/3) V_0/M_P^2$ to obtain the second expression. 
where $T_R$ is the reheat temperature after inflation. 
Assuming instant reheating, the reheat temperature is given by $T_R \sim \sqrt{H_{\rm inf} M_P}$. 
To satisfy the TCC bound, $H_{\rm inf} \lesssim 1$ GeV and thus $T_R \lesssim 10^8$ GeV. 
As a benchmark, let us take $H_{\rm inf} = 0.1$ GeV and $N=40$. 
In this case,  the running of the spectral index is  predicted to be $\alpha \simeq -0.0062$ \cite{Okada:2016ssd}. 
This is consistent with the Planck 2018 result $\alpha = - 0.0045\pm 0.0067$ \cite{Akrami:2018odb}
  and can be further tested in the future \cite{Abazajian:2013vfg}.

Next we identify the inflaton in the IPI scenario with the $B-L$ Higgs field, $\phi = \sqrt{2}  {\rm Re}[\Phi]$. 
In our inflation analysis, we employ the Renormalization Group (RG) improved effective inflaton potential given by 
\bea
V( \phi) =  \lambda_\phi (\phi) \left( \Phi^\dagger \Phi - \frac{v_{BL}^2}{2}  \right)^2 \simeq \frac{1}{4} \lambda_\phi ( \phi)\; \phi^4. 
\label{eq:VEff1}
\eea
Here, we have used $ \phi \gg v_{BL}$ during the inflation, 
  and $\lambda_\phi (\phi)$ is the solution to the following RG equations:  
\bea
 \phi  \frac{d g}{d  \phi} &=& \frac{12}{16 \pi^2}  g^3,         \nonumber\\
 \phi \frac{d Y_{i}}{d  \phi}   &=& \frac{1}{16 \pi^2} Y_i \, \left(Y_i^2+\frac{1}{2} \sum_{j=1}^3 Y_j^2-6 g^2 \right) ,   
\nonumber\\
 \phi \frac{d \lambda_\phi}{d  \phi}  &=& \beta_{\lambda_\phi}. 
  \label{eq:RGEs}
\eea
The beta-function of $\lambda_\phi$ is approximately given by 
\bea
%\beta_{\lambda_\phi} \!=\! \frac{1}{16 \pi^2}\!\! \left(\!20 \lambda_\phi^2  \!- 48\lambda_\phi  g^2\!+2 \lambda_\phi\sum_i Y_i^2 
%\right. 
%\nonumber \\
%&& \left.
\beta_{\lambda_\phi}  \simeq   \frac{1}{16 \pi^2} 
\left( 96 g^4 -\! \sum_{i=1}^3 Y_i^4 \right)   
\label{eq:BGen}
\eea
for $\lambda_\phi \ll g^4, Y_i^4$. 
We later justify the validity of this approximation.

The RG improved effective potential in Eq.~(\ref{eq:VEff1}) leads to
\bea
\frac{V_1}{M^3}&=& \left.\frac{1}{4} (4 \lambda_\phi + \beta_{\lambda_\phi})\right|_{ \phi= M},\nonumber \\
\frac{V_2}{M^2}&=&  \left.\frac{1}{4} (12\lambda_\phi + 7\beta_{\lambda_\phi}+M \beta_{\lambda_\phi}^\prime)\right|_{ \phi= M}, \nonumber \\
\frac{V_3}{M}&=&  \left.\frac{1}{4} (24\lambda_\phi \!+\! 26\beta_{\lambda_\phi}\!+\!10M \beta_{\lambda_\phi}^\prime \!\! +\!M^2 \beta_{\lambda_\phi}^{\prime\prime})\right|_{ \phi= M}, 
\label{eq:ICons2}
\eea
where the prime denotes a derivative with respect to the field $\phi$.
Since $M$ is approximately an inflection-point, $V_1/M^3\simeq 0$ and $V_2/M^2\simeq 0$, 
   which lead to $\beta_{\lambda_\phi} (M)\simeq -4\lambda_\phi(M)$ 
   and $M\beta_{\lambda_\phi}^{\prime}(M) \simeq  16 \lambda_\phi (M)$. 
As we will show later, these equations requires $g(M)$ and $Y_3 (M)$ to be of the same order (assuming $Y_1 (M) \simeq Y_2 (M) \ll  Y_3 (M)$, for simplicity)  and $\lambda (M) \propto g (M)^6$. 
We can approximate $M^2 \beta_{\lambda_\phi}^{\prime\prime}(M) = - M \beta_{\lambda_\phi}^{\prime}(M)  +\phi \frac{{ d} }{{d}\phi} ( \phi \beta_{\lambda_\phi}^{\prime} ) |_{\phi = M} \simeq - M \beta_{\lambda_\phi}^{\prime}(M) \simeq -16 \lambda_\phi(M)$. 
Here, we have neglected the $\phi \frac{{ d} }{{d}\phi} ( \phi \beta_{\lambda_\phi}^{\prime} ) |_{\phi = M}$ term which is a polynomial of degree eight in terms of $g (M)$ and $Y_3 (M)$, and  $M \beta_{\lambda_\phi}^{\prime}(M)$ is a polynomial of degree six.  
Hence, the last equation in Eq.~(\ref{eq:ICons2}) is simplified as $V_3/M \simeq 16 \lambda_\phi(M)$. 
Comparison with $V_3/M$ in Eq.~(\ref{eq:FEq-V1V2}) with $V_0\simeq (1/4) \lambda_\phi(M) M^4$ yields %the final expression for $\lambda_\phi$ at the inflation scale $\phi = M$, 
\bea
\lambda_\phi(M)\simeq 2.31 \times 10^{-15} \left(\frac{M}{M_{P}}\right)^2. 
\label{eq:FEq1} 
\eea
Using this expression of $\lambda_\phi(M)$, we find $H_{\rm inf}$ as a function of $M$:
\bea
%H_{\rm inf} &\simeq& \sqrt{\frac{V_0}{3 M_P^2}} \simeq 3.45\times 10^{10} \;{\rm GeV} \;\left(\frac{M}{M_P}\right)^3, 
%\nonumber \\
%r &\simeq& 1.86 \times 10^{-8}  \left(\frac{M}{M_{P}}\right)^6,
%
H_{\rm inf}  \simeq  \sqrt{\frac{V_0}{3 M_P^2}} \simeq 3.38\times 10^{10} \;{\rm GeV} \;\left(\frac{M}{M_P}\right)^3. 
\label{eq:FEqR} 
\eea
For our benchmark $H_{inf} = 0.1$ GeV to satisfy the TCC bound, 
  we find $M = 3.50 \times 10^{14}$ GeV. 
In this case, the prediction for $r$ is tiny, namely, $r \sim 10^{-31}$.

Let us consider some low energy predictions of the IPI scenario. 
In the following analysis, we set $Y_1 \simeq Y_2 \ll Y_3 \equiv Y$. 
The inflection point condition, $\beta_{\lambda_\phi}(M) \simeq 0$, leads to 
\bea
Y(M)\simeq 3.13\;g(M). 
\label{eq:FEq3}
\eea 
With this relation, the RG equations in Eq.~(\ref{eq:RGEs}), and another inflection point condition, 
   $M\beta_{\lambda_\phi}^{\prime}(M) \simeq  16 \lambda_\phi (M)$, 
   we obtain 
\bea
\lambda_\phi(M) \simeq 3.09 \times 10^{-3} \, g(M)^6. 
\eea
Comparing this with Eq.~(\ref{eq:FEq1}), 
the $B-L$ gauge coupling is expressed as a function of $M$: 
\bea
g(M)\simeq 9.52 \times 10^{-3}  
 \left(\frac{M}{M_{P}}\right)^{1/3}.
\label{eq:FEq2} 
\eea 
Let us now evaluate the low-energy values of  $g(\varphi)$,  $Y(\varphi)$ and $\lambda (\varphi)$. 
Since $g(M), Y(M) \ll1$, their low energy value can be approximated as 
\bea
g(\phi)   &\simeq& g(M) +\beta_{g}(M) \ln \left[\frac{\phi}{M}\right]  ,\nonumber \\ 
Y(\phi)  &\simeq& Y(M)+ \beta_{Y}(M) \ln \left[\frac{\phi}{M}\right], 
\label{eq:gYatVEV}
\eea
where $\beta_{g}(M)$ and $\beta_{Y}(M)$, respectively, are the  beta-functions for $g$ and $Y$ in Eq.~(\ref{eq:RGEs}). 
Plugging these into the expression for $\beta_{\lambda}(\phi)$ in Eq.~(\ref{eq:BGen}) we obtain 
\bea
 16\pi^2 \beta_{\lambda_\phi} &\simeq& 16 \lambda_\phi (M) \ln \left[\frac{\phi}{M}\right], 
\label{eq:betaL}
\eea
where we have used the approximations $\beta_{\lambda_\phi} (M)  \simeq 0$ and $M\beta_{\lambda_\phi}^{\prime}(M) \simeq  16 \lambda_\phi (M)$. 
Then, $\lambda_\phi$ value can be estimated by solving its RG equation in Eq.~(\ref{eq:RGEs}),
\bea
\lambda_\phi( \phi) &\simeq& \lambda_\phi (M) + \frac{1}{2 \pi^2} \lambda_\phi (M) \left(\!\ln \left[\frac{ \phi}{M}\right] \!\right)^2,  
\nonumber \\
 &\simeq & 1.85\times 10^{-14}\left(\!\frac{M}{M_P}\!\right)^2\left(\!\ln \left[\frac{ \phi}{M}\right] \!\right)^2, 
\label{eq:Lmu} 
\eea
for $\phi \ll M$. 
The particle mass spectrum (see Eq.~(\ref{eq:mass})) is found to be 
\bea
m_\phi &\simeq& 1.92 \times 10^{-7} \, v_{BL} \left({\rm ln}\left[\frac{M}{v_{BL}}\right] \right)\left(\frac{M}{M_P}\right), 
\nonumber \\
m_{Z^\prime} &\simeq& 1.91 \times 10^{-2} \, v_{BL}
 \left(\frac{M}{M_{P}}\right)^{\!\!1/3}
\nonumber \\
m_{N^{3}} &\simeq& 1.11 \, m_{Z^\prime}. 
\label{eq:ratiophiz} 
\eea 
For our benchmark $H_{inf} = 0.1$ GeV ($M = 3.47 \times 10^{14}$ GeV), 
  let us fix $v_{BL} = 3.06\times 10^{12}$ GeV for the rest of our analysis, 
  so that $m_\phi \simeq 400$ GeV, 
$m_{Z^\prime} \simeq 3.05 \times10^9$ GeV, 
and $m_{N^3} \simeq  3.38 \times 10^9$ GeV.

%%%%%%%%%%%%%%%%%%%%%%%%%%%%%%%%
\begin{figure}[t]
\begin{center}
\includegraphics[scale=0.6]{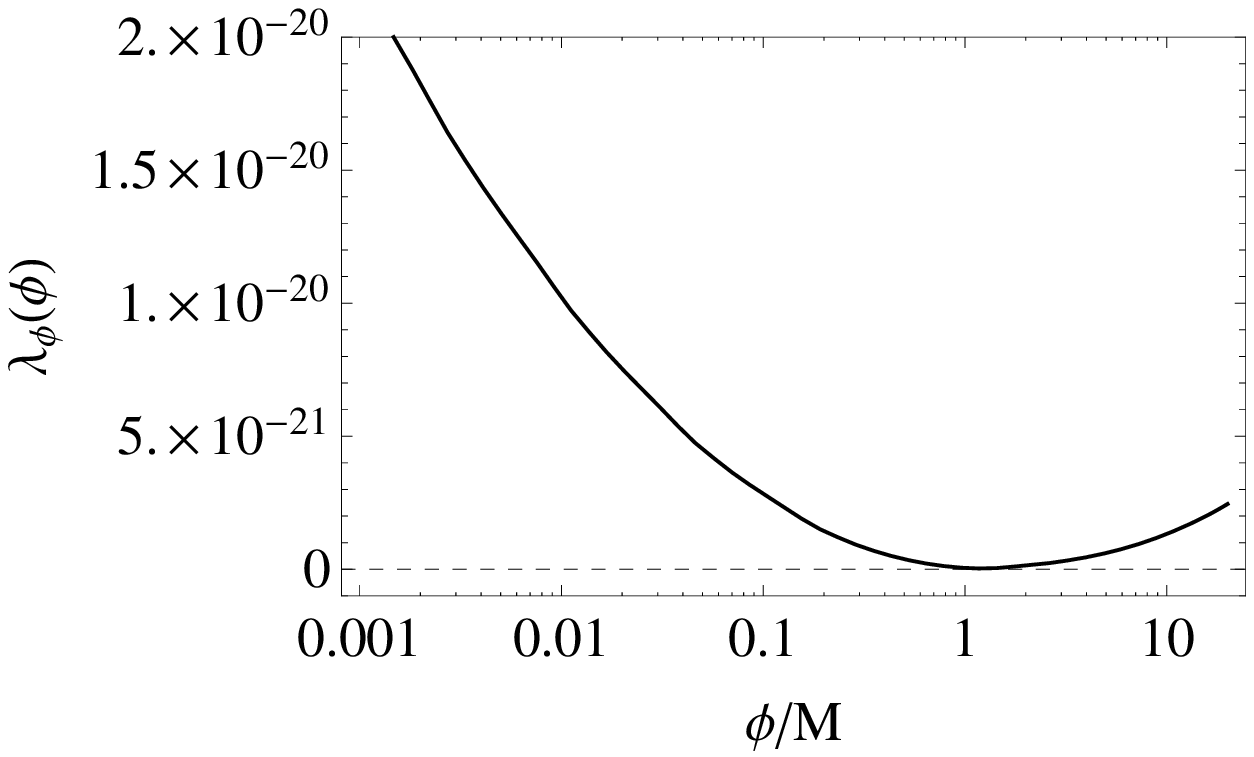} \;
\includegraphics[scale=0.6]{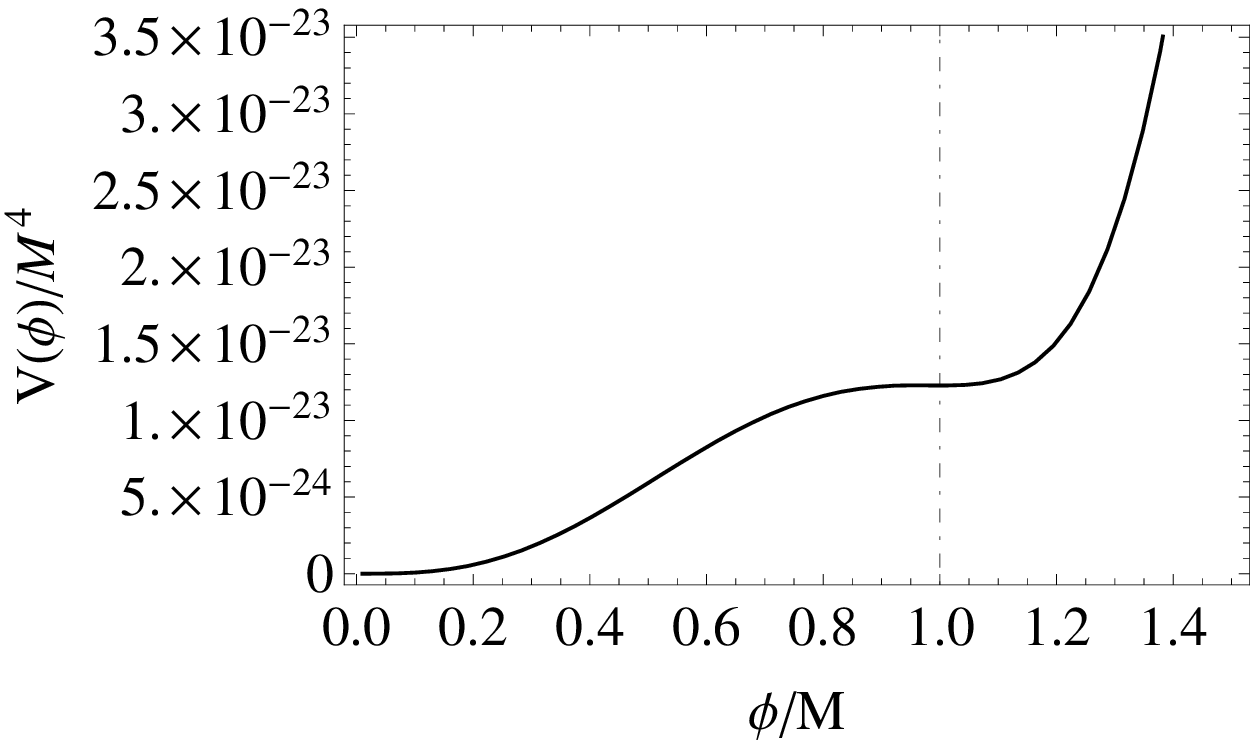}       
\end{center}
\caption{
For $M = 3.47 \times 10^{14}$ GeV, the top panel shows the RG running of inflaton quartic coupling as a function of $ \phi/M$, 
  along with the dashed horizontal line corresponding to $\lambda_\phi=0$. 
The bottom panel shows the RG improved effective inflaton potential with an approximate inflection-point 
  at $ \phi \simeq M$ (indicated by vertical dashed-dotted line). 
}
\label{fig:InfPot}
\end{figure}
%%%%%%%%%%%%%%%%%%%%%%%%%%%%%%%%%%

With $M \simeq 3.47 \times 10^{14}$ GeV, 
  we show in Fig.~\ref{fig:InfPot} the RG running of  the inflaton quartic coupling (top), 
  and the corresponding RG improved effective inflaton potential (bottom). 
%Corresponding values of $g (M)= 5.01 \times 10^{-4}$, $Y (M) \simeq 1.56 \times 10^{-3}$, and $\lambda_\phi (M)  \simeq 9.70 \times10^{-24}$. 
The top panel shows that the RG running of the quartic coupling (solid curve) exhibits a minimum at $ \phi \simeq M$ 
   with an almost vanishing value, or equivalently, 
   $\lambda_\phi (M)\simeq 0$ and $\beta_{\lambda_\phi} (M) \simeq 0$.\footnote{Similar conditions have been obtained in Ref.~\cite{Okada:2015lia} to stabilize the inflaton potential.} 	
This behavior of the running of $\lambda_\phi$ is essential to realize an approximate inflection-point at $\phi = M$
   as shown in the bottom panel.

%%%%%%%%%%%%%%%%%%%%%%%
\section{Reheating and resonant leptogenesis} 
\label{sec:4}
%%%%%%%%%%%%%%%%%%%%%%%
After inflation, the universe is thermalized by the SM particles created from the inflaton decay. 
To evaluate the reheat temperature, we consider mixed  quartic couplings of the inflaton
  with the SM Higgs doublets: 
\bea
V \supset  4 \lambda^\prime  \left(\Phi^\dagger \Phi\right) \left(H_u^\dagger H_u + H_d^\dagger H_d\right)
\supset 
 \lambda^\prime v_{BL}  \phi  h^2,   
\label{eq:InfPot}
\eea 
where we have introduce a common coupling $\lambda^\prime$, 
   and $h$ is the SM-like Higgs boson. 
The decay width of the inflaton is given by  
\bea
\Gamma ( \phi \to h h ) \simeq  \frac{{\lambda^{\prime}}^2 v_{BL}^2 }{4 \pi \, m_ \phi},   
\label{eq:gamma2}
\eea
and reheat temperature is estimated to be 
\begin{eqnarray}
    T_R &\simeq& \left(\frac{90}{\pi^2 g_*}\right)^{1/4} \sqrt{\Gamma M_P}  \nonumber\\
    &\simeq & 10^{8} \;{\rm GeV} \left(\frac{\lambda^\prime}{2.70 \times 10^{-12}}\right), 
\label{Lambda}
\end{eqnarray} 
where $g_* \simeq 100$.   
In order for this $T_R$ value not to exceed the maximum reheat temperature 
  evaluated from instant reheating, $T_R \sim 10^8$ GeV, 
  we require $\lambda^\prime < 2.70 \times 10^{-12}$. 
We can easily verify that the $\lambda^\prime$ value is sufficiently small to be neglected in our previous analysis.

In models with the type-I seesaw mechanism, leptogenesis \cite{Fukugita:1986hr} is the simplest mechanism
  for generating the observed baryon asymmetry in the universe. 
For thermal leptogenesis with hierarchical RHN mass spectrum, there is a lower bound on 
  the lightest RHN mass of around $10^{9-10}$ GeV \cite{Buchmuller:2002rq} 
  and thus the reheat temperature must be higher than this value. 
Since the maximum $T_R \sim 10^8$ GeV is lower than this value, 
  successful leptogenesis requires an enhancement of the CP-asymmetric parameter 
  through a degenerate RHN mass spectrum, namely, resonant leptogenesis \cite{Flanz:1996fb, Pilaftsis:1997jf}.  
For the resonant leptogenesis, we set $Y_1 \simeq Y_2$ for the Majorana Yukawa coupling of two RHNs.

Since the RHNs have $B-L$ gauge interaction, 
  they can stay in thermal equilibrium with the plasma of the SM particles. 
As a result, the generation of lepton asymmetry is suppressed 
  until the $B-L$ interaction is frozen \cite{Iso:2010mv}. 
We now derive the condition for successful leptogenesis. 
Let us consider the process 
    ${N_R}^{1,2} N_R^{1,2} \leftrightarrow  Z^\prime \to \overline{f_{SM}} f_{SM}$, where $f_{SM}$ are the SM fermions. 
We require this process to decouple before the temperature of the universe drops to 
    $T \sim m_{N^1} \simeq m_{N^2} \equiv M_N$.  
Since $M_N \ll m_{Z^\prime}$ in our setup, the $Z^\prime$ mediated process is effectively 
  described as a four-Fermi interaction, and its thermal-averaged cross section can be approximated as \cite{Okada:2016ssd}
\bea
\langle \sigma v \rangle \simeq \frac{13}{192\pi} \frac{T^2}{v_{BL}^4},   
\eea 
   for temperature $T \gtrsim M_N$. 
In order for this process to be decoupled at $T \sim M_N$, we impose $\Gamma/H(T) < 1$, 
   where $\Gamma = n_{eq}(T) \langle  \sigma v \rangle$ is the annihilation/creation rate of the RHNs
   with equilibrium number density $n_{eq}(T) \simeq 2 T^3/\pi^2$,
   and the Hubble parameter $H(T) \sim T^2/M_P$. 
This leads to a lower bound on $v_{BL}$: 
\bea
v_{BL} > 10^{9} \;{\rm GeV} \left(\frac{M_{N}}{1.1\times 10^{7} \; {\rm GeV}}\right)^{3/4}. 
 \label{cond1}
\eea

There is another process which can suppress the generation of lepton asymmetry, 
  namely, ${N_R}^{1,2} N_R^{1,2} \leftrightarrow  \phi \phi$ 
  when $M_N > m_\phi$ \cite{Dev:2017xry}. 
The thermal-averaged cross section of this process is roughly given by \cite{Plumacher:1996kc}
\bea
   \langle \sigma v \rangle \simeq \frac{1}{4 \pi}   \frac{M_N^2}{v_{BL}^4}. 
\eea
Requiring $\Gamma/H(T) < 1$ at $T=M_N$, we obtain 
\bea
 M_N < (2 \pi^3)^{1/3}  \, v_{BL} \, \left( \frac{v_{BL}}{M_P} \right)^{1/3}. 
\label{cond2}
\eea
We can see that the conditions in Eqs.~(\ref{cond1}) and (\ref{cond2}) are satisfied with 
  our benchmark choice of $v_{BL} =  3.06 \times 10^{12}$ GeV and 
   $M_{N} \ll  m_{Z^\prime} \sim m_{N^3} \sim 3 \times 10^9$ GeV. 
Therefore, the observed baryon asymmetry in the universe can be reproduced by resonant leptogenesis.

%%%%%%%%%%%%%%%%%%%%%%%%%%%%%%%
\section{Axion dark matter with $H_{\rm inf} \lesssim 0.1\, {\rm GeV}$}
\label{sec:5}
%%%%%%%%%%%%%%%%%%%%%%%%%%%%%%%
The axion scenario not only offers an elegant solution to the strong CP problem
   but also provides us with a compelling DM candidate in the form of axion. 
However, this scenario in general suffers from cosmological problems, 
   such as the domain wall and the isocurvature problems. 
For a review, see, for example, Ref.~\cite{Kawasaki:2013ae}. 
The domain wall problem arises because topological defects (strings and domain walls) associated with PQ symmetry breaking evolve to dominate the energy density of the universe which is inconsistent with cosmological observation. 
It can be solved if inflation takes places after the PQ symmetry breaking\footnote{For a resolution of the domain wall problem without inflation, see Ref.~\cite{Kibble:1982ae}.}, 
  namely, $H_{\rm inf} < F_a=v_{PQ}/N_{DM}$, 
  where $F_a$ is the axion decay constant and $N_{DM}$ ($N_{DW} \neq 1$ in general) is the domain wall number. 
In our case, $N_{DW} = 12$.

The measurement of supernova SN1987A pulse duration provides a model-independent constraint
 on the axion decay constant $F_a  \gtrsim 4 \times 10^8$ GeV \cite{Raffelt:2006cw}.  
On the other hand, if inflation takes place after the PQ symmetry breaking, 
   the axion obtains large fluctuations that generates  isocurvature density perturbations, 
   which are severely constrained by the Planck measurements \cite{Akrami:2018odb}.  
With a natural assumption $\theta_a ={\cal O}(1)$ for the initial displacement of the axion field (misalignment angle) from the potential minimum at the onset of oscillations, 
we obtain an upper bound on $H_{\rm inf}$  \cite{Kawasaki:2013ae}  
\bea
H_{\rm inf} < 2.08 \times 10^7 \;  {\rm GeV} \left(\frac{F_a}{7.11 \times 10^{11} {\rm GeV}}\right)^{0.405}.   
\label{eq:Hiso1}
\eea
The Hubble scale to satisfy the TCC bound ($H_{\rm inf} \lesssim 0.1$ GeV) is fully compatible with this bound, 
  and our model is therefore free from the axion cosmological problems.

After the QCD phase transition, the coherently oscillating axion field behaves like cold DM 
  whose abundance is given by \cite{Kawasaki:2013ae}  
\bea
\Omega_a h^2 &\simeq&  0.12 \; 
\left[ \theta_a^2 + \left(\frac{H_{\rm inf}}{2\pi F_a}\right)^2\right]\; 
 \left(\frac{F_a}{7.11 \times 10^{11}\; {\rm GeV}}\right)^{1.19}
\nonumber \\ 
&\simeq& 0.12\;  \theta_a^2 \left(\frac{F_a}{7.11 \times 10^{11}\; {\rm GeV}}\right)^{1.19}, 
\label{eq:DMab}
\eea
where we have used $H_{\rm inf}/(2\pi F_a) \ll 1$ to obtain the final expression. 
To reproduce the observed DM abundance of $\Omega_{a} h^2 = 0.12$ \cite{Aghanim:2018eyx}, 
   we set $\theta_a \simeq 1$ (natural choice) and therefore $F_a = 7.04 \times 10^{11}$ GeV. 
Axion DM scenario with larger $F_a$ values requires $\theta_a < 1$.\footnote{Recently it has been pointed out in Refs.~\cite{Graham:2018jyp, Guth:2018hsa} that an eternal inflation scenario with $H_{\rm inf} < \Lambda_{QCD} \simeq 0.1$ GeV dynamically realizes misalignment angle $\theta_a\ll 1$.}

\begin{comment}
However, recently it has been pointed out in Refs.~\cite{Graham:2018jyp, Guth:2018hsa} that 
  for $H_{\rm inf} \lesssim \Lambda_{QCD} \simeq 0.1$ GeV 
  there is an upper bound on the misalignment angle \cite{Guth:2018hsa}: 
\bea
\theta_a \lesssim 0.34 \left(\frac{H_{\rm inf}}{0.1 \, {\rm GeV}}\right)^2.  
\label{eq:theta}
\eea
In Ref.~\cite{Guth:2018hsa}, the axion abundance is numerically evaluated
  for the maximum value of $\theta_a$ to satisfy Eq.~(\ref{eq:theta}) 
  and the relation between $F_a$ and $H_{\rm inf}$ to reproduce the observed DM abundance $\Omega_a h^2 = 0.12$ 
  has been obtained:   
\bea
F_a \simeq 6.01 \times 10^{12} \, {\rm GeV} \left(\frac{0.1 \, {\rm GeV}} {H_{\rm inf}}\right)^{3.36}. 
\eea
which is applicable for $H_{\rm inf} < 172$ MeV. 
Therefore, the upper bound on $F_a$ is significantly relaxed for  low-scale inflation. 
Satisfying the TCC bound on $H_{\rm inf} \lesssim 0.1$ GeV, 
 the axion DM scenario in our model remains viable with $F_a$ larger than $10^{12}$ GeV. 
\end{comment}

%%%%%%%%%%%%%%%%%%%%%%%%%%%%%%%%%%%%%%%%%
\section{Conclusions}
\label{sec:6}
%%%%%%%%%%%%%%%%%%%%%%%%%%%%%%%%%%%%%%%%
Inspired by the Trans-Planckian Censorship Conjecture, 
  we have presented a well motivated U(1)$_{B-L} \times$U(1)$_{PQ}$ extension of the Standard Model. 
The $B-L$ component allows us to incorporate inflection-point inflation scenario at low scale 
  that is compatible with the TCC bound on $H_{\rm inf} \lesssim 1$ GeV.  
Its inflationary predictions are  consistent with the Planck 2018 measurements, 
and the prediction for the running of spectral index of $\alpha  \simeq -0.0062$ 
  which can be tested in the future. 
Because of the low scale inflation, the axion dark matter scenario is free from the domain wall and isocurvature  problems. 
The seesaw mechanism is automatically incorporated in the model and 
   the observed baryon asymmetry of the universe can be reproduced via resonant leptogenesis.

%%%%%%%%%%%%%%%%%%%%%%%%%%%%%%%%%%%%%%%%%
\section*{Acknowledgements}
%%%%%%%%%%%%%%%%%%%%%%%%%%%%%%%%%%%%%%%%%
This work is supported in part by the United States Department of Energy grant DE-SC0012447 (N.~O) 
and DE-SC0013880 (D.~R and Q.~S).

%%%%%%%%%%%%%%

%%%%%%%%%%%%%%%%%%%%%%%%%%%%%%%%%%%%%%

\end{document}